\documentclass[12pt,letterpaper]{article}

\usepackage{verbatim}
\usepackage{amsmath}
\usepackage{graphicx}
\usepackage{cite}
\usepackage{subfig}
\usepackage{setspace}

\def\Re{{\cal R \mskip-4mu \lower.1ex \hbox{\it e}\,}}
\def\Im{{\cal I \mskip-5mu \lower.1ex \hbox{\it m}\,}}
\def\ie{{\it i.e.}}
\def\eg{{\it e.g.}}

\def\sub#1{_{\lower.25ex\hbox{$\scriptstyle#1$}}}

\def\to{\rightarrow}

\def\subw{_{\rm w}}
\def\mh{\ifmmode m\sbl H \else $m\sbl H$\fi}
\def\mch{\ifmmode m_{H^\pm} \else $m_{H^\pm}$\fi}
\def\mt{\ifmmode m_t\else $m_t$\fi}
\def\mc{\ifmmode m_c\else $m_c$\fi}
\def\mz{\ifmmode M_Z\else $M_Z$\fi}
\def\mw{\ifmmode M_W\else $M_W$\fi}
\def\mws{\ifmmode M_W^2 \else $M_W^2$\fi}
\def\mhs{\ifmmode m_H^2 \else $m_H^2$\fi}   
\def\mzs{\ifmmode M_Z^2 \else $M_Z^2$\fi}
\def\mts{\ifmmode m_t^2 \else $m_t^2$\fi}
\def\mcs{\ifmmode m_c^2 \else $m_c^2$\fi}
\def\mchs{\ifmmode m_{H^\pm}^2 \else $m_{H^\pm}^2$\fi}
\def\ztwo{\ifmmode Z_2\else $Z_2$\fi}
\def\zone{\ifmmode Z_1\else $Z_1$\fi}
\def\mtwo{\ifmmode M_2\else $M_2$\fi}
\def\mone{\ifmmode M_1\else $M_1$\fi}
\def\tb{\ifmmode \tan\beta \else $\tan\beta$\fi}
\def\xw{\ifmmode x\subw\else $x\subw$\fi}
\def\ch{\ifmmode H^\pm \else $H^\pm$\fi}
\def\lum{\ifmmode {\cal L}\else ${\cal L}$\fi}
\def\inpb{\ifmmode {\rm pb}^{-1}\else ${\rm pb}^{-1}$\fi}
\def\infb{\ifmmode {\rm fb}^{-1}\else ${\rm fb}^{-1}$\fi}
\def\epem{\ifmmode e^+e^-\else $e^+e^-$\fi}
\def\ppb{\ifmmode \bar pp\else $\bar pp$\fi}
\def\pbp{\ifmmode ~^(\bar p^)p\else $~^(\bar p^)p$\fi}
\def\bsg{\ifmmode B\to X_s\gamma\else $B\to latexilaX_s\gamma$\fi}
\def\bsll{\ifmmode B\to X_s\ell^+\ell^-\else $B\to X_s\ell^+\ell^-$\fi}
\def\bstt{\ifmmode B\to X_s\tau^+\tau^-\else $B\to X_s\tau^+\tau^-$\fi}

\newskip\zatskip \zatskip=0pt plus0pt minus0pt
\def\matth{\mathsurround=0pt}

\def\atversim#1#2{\lower0.7ex\vbox{\baselineskip\zatskip\lineskip\zatskip
  \lineskiplimit 0pt\ialign{$\matth#1\hfil##\hfil$\crcr#2\crcr\sim\crcr}}}

\renewcommand{\thefootnote}{\fnsymbol{footnote}}

\begin{document} \begin{titlepage} 
\rightline{\vbox{\halign{&#\hfil\cr
&SLAC-PUB-14544\cr
}}}
\vspace{1in} 
\begin{center}

{{\Large\bf Zeroing in on Supersymmetric Radiation Amplitude Zeros}
\footnote{Work supported by the Department of 
Energy, Contract DE-AC02-76SF00515}\\}
\medskip
\medskip
\normalsize 
{\large JoAnne L. Hewett, Ahmed Ismail, and Thomas G. Rizzo \\
\vskip .6cm
SLAC National Accelerator Laboratory,  \\
2575 Sand Hill Rd, Menlo Park, CA 94025, USA\\}
\vskip .5cm

\end{center} 
\vskip 0.8cm

\begin{abstract} 

Radiation amplitude zeros have long been used to test the Standard Model. Here, we consider the supersymmetric radiation amplitude zero in chargino-neutralino associated production, which can be observed at the luminosity upgraded LHC. Such an amplitude zero only occurs if the neutralino has a large wino fraction and hence this observable can be used to determine the neutralino eigenstate content.  We find that this observable can be measured by comparing the $p_T$ spectrum of the softest lepton in the trilepton $\tilde{\chi}_1^\pm \tilde{\chi}_2^0$ decay channel to that of a control process such as $\tilde{\chi}_1^+ \tilde{\chi}_1^-$ or $\tilde{\chi}_2^0 \tilde{\chi}_2^0$. We test this technique on a previously generated model sample of the 19 dimensional parameter space of the phenomenological MSSM, and find that it is effective in determining the wino content of the neutralino.

\end{abstract} 
\vspace{1in}
\begin{center}
{\textit{Dedicated to the memory of Ulrich Baur}}
\end{center}

\renewcommand{\thefootnote}{\arabic{footnote}} \end{titlepage}

% \doublespacing

\section{Introduction}

As the first data from the Large Hadron Collider (LHC) arrive, supersymmetry remains one of the leading candidates for physics beyond the Standard Model (SM), providing a resolution for the hierarchy problem, a mechanism for unification of the forces, as well as a natural dark matter candidate. The Minimal Supersymmetric Standard Model (MSSM) is the simplest extension of the SM that incorporates supersymmetry, since it has the minimal number of additional particles, and has been well studied for decades in the literature~\cite{Martin:1997ns}. Nonetheless, the breaking of supersymmetry, necessitated by the lack of observation of super-partners, introduces $\sim 105$ new parameters, even in its minimal form. Models of supersymmetry breaking can reduce the size of this parameter set by assuming theoretical relations at high energy scales.

If supersymmetry is found to describe physics at and beyond the electroweak scale, much work will be needed to extract the various parameters of the theory, particularly if no specific breaking scenario such as minimal supergravity (mSUGRA)~\cite{Chamseddine:1982jx} is realized in nature. In particular, knowledge of the eigenstate content of the electroweak gaugino sector will be important to test supersymmetry and probe physics at even higher scales, as well as to provide valuable information for the study of dark matter annihilation. Recently, a method has been proposed to determine at the LHC the content of the lightest neutralino in the case that it is the lightest supersymmetric particle (LSP)~\cite{Kane:2011tv}. However, techniques to decipher the rest of the neutralino sector in a hadron collider environment remain relatively unexplored. Such tests of the electroweak gaugino sector are easily performed at a high energy $e^+ e^-$ collider~\cite{Feng:1995zd}.

Radiation amplitude zeros (RAZ), the vanishing of amplitudes of certain processes with external photons in specific kinematic regions, have been used to test the structure of the Standard Model gauge sector and search for anomalous couplings of the $W$ boson~\cite{Baur:1994sa,:2008vja,Chatrchyan:2011rr}. After the existence of such an amplitude zero was first noticed in the computation of $W \gamma$ production~\cite{Brown:1979ux}, a general theory of RAZ was developed~\cite{Zhu:1980sz,Brodsky:1982sh}, explaining the phenomenon as a result of gauge and Lorentz invariance. Discoveries of RAZ in other SM processes, as well as various extensions~\cite{Brown:1984nj}, soon followed, with zeros predicted in $WZ$ production~\cite{Baur:1994ia}, quark scattering with an associated photon~\cite{Brodsky:1982sh}, charged Higgs decay~\cite{Mukhopadhyaya:1990ba}, leptoquark production~\cite{Deshpande:1994vf,Doncheski:1998cv}, and $\tilde{W} \tilde{\gamma}$ production in exact supersymmetry~\cite{Barger:1983wc}.

Here, we consider the potential RAZ in associated chargino-neutralino production in the MSSM, and investigate how it can be used to constrain the content of the neutralino. We find that an observable RAZ only occurs for a wino-like neutralino, and so the wino content of the neutralino may be probed by searching for an amplitude zero in this associated production channel. This zero is observable in the trilepton decay channel, by examining the $p_T$ distribution of the softest lepton. As we will see below, due to the RAZ, the $p_T$ spectrum of the softest lepton in the trilepton final state tends to drop off faster with increasing $p_T$ when the neutralino is wino-like. By comparing chargino-neutralino production to other supersymmetric processes that don't exhibit a RAZ, we can place a limit on the relevant neutralino mixing matrix elements. While we consider $\tilde{\chi}_1^\pm \tilde{\chi}_2^0$ production, our method is limited only by statistics, and could in principle be used to test the wino content of the heavier neutralinos $\tilde{\chi}_3^0$ and $\tilde{\chi}_4^0$ if they are kinematically accessible. We assume that the $\tilde{\chi}_1^0$ is the LSP throughout, and its lack of a detector signature renders our technique ineffective for determining the $\tilde{\chi}_1^0$  wino content. By considering trilepton events at the upgraded LHC with high luminosity, we demonstrate the ability to constrain the neutralino mixing matrix.

To evaluate the potential of our technique, we make use of a previous scan~\cite{Berger:2008cq} over the phenomenological MSSM (pMSSM) parameter space and examine the signature of amplitude zeros in a sample of supersymmetric models. This scan resulted in $\sim 7 \cdot 10^4$ visible models, and hence provides an excellent testing ground for observing a supersymmetric RAZ. Approaches such as the pMSSM do not involve theoretical assumptions at the high scale, and while the resulting parameter space is significantly larger, results derived from this approach are more general in scope. The pMSSM imposes only phenomenologically motivated constraints on the MSSM, and has the advantage of having more predictive power than the full MSSM with negligible loss of generality. 

This paper is organized as follows. In Sections~\ref{pmssm} and~\ref{raz}, we review the pMSSM and the theory of RAZ. Then, we present our application of RAZ to the pMSSM to determine the wino content of the neutralino in Section~\ref{analysis}. Section~\ref{conclusion} contains our conclusions.

\section{The pMSSM}
\label{pmssm}

In contrast to models which reduce the number of free parameters in the MSSM by making theoretically motivated assumptions, \eg, universal soft supersymmetry breaking terms or gauge coupling unification, the pMSSM only imposes constraints that are driven by experiment. While there are more parameters in the pMSSM than in models which assume a specific supersymmetry breaking mechanism, it provides a more general framework in which to conduct our analysis. We now briefly summarize the assumptions used in generating the pMSSM model sample. This sample is limited to the CP conserving MSSM with Minimal Flavor Violation. $\Delta F = 2$ flavor changing neutral currents limit the sfermion mass matrices to be essentially diagonal, with the first two generations degenerate. Also, strong constraints on CP violation suggest that the soft breaking terms conserve CP, eliminating many complex phases that are allowed in the full MSSM. Finally, the trilinear A-term couplings for the first two generations do not have significant experimental consequences, since they are proportional to the fermion masses, and so they are neglected. Together, these constraints leave only 19 free parameters in the pMSSM:

\begin{itemize}
\item $M_{1,2,3}$, the gaugino masses
\item $\mu$, the Higgs(ino) mass parameter
\item $\tan \beta$, the ratio of the Higgs doublet vacuum expectation values
\item $m_A$, the mass of the CP-odd Higgs boson
\item Five sfermion masses for the first two generations
\item Five sfermion masses for the third generation
\item $A_{b,t,\tau}$, the third generation trilinear couplings.
\end{itemize}

A scan over this parameter space has been previously performed~\cite{Berger:2008cq}, where the parameters above were allowed to vary within the following ranges that were chosen in the interest of producing observable signals at the LHC:

\begin{align}
50 \mbox{ GeV} \le |M_{1,2}, \mu| \le& 1 \mbox{ TeV}, \nonumber \\
100 \mbox{ GeV} \le M_3 \le& 1 \mbox{ TeV}, \nonumber \\
1 \le \tan \beta \le& 50, \\
43.5 \mbox{ GeV} \le m_A \le& 1 \mbox{ TeV}, \nonumber \\
100 \mbox{ GeV} \le m_{\tilde{f}} \le& 1 \mbox{ TeV}, \nonumber \\
|A_{b,t,\tau}| \le& 1 \mbox{ TeV}. \nonumber
\end{align}

Models with tachyons, unbounded scalar potentials, or scalar potential minima that violate charge or color conservation were rejected. In addition, the lightest supersymmetric particle (LSP) was required to be the lightest neutralino $\tilde{\chi}_1^0$ and a thermal relic. After sampling $10^7$ points from the parameter region above using flat priors, these models were subjected to numerous experimental constraints, including $\Delta \rho$, $b \rightarrow s \gamma$, $B_s \rightarrow \mu^+ \mu^-$, $B \rightarrow \tau \nu$, $(g - 2)_\mu$, the WMAP relic density measurement, direct dark matter detection bounds, and direct collider searches for sparticles and Higgs bosons at LEP and the Tevatron. A complete discussion of the constraints applied may be found in~\cite{Berger:2008cq}. Of the $10^7$ sample models, $\sim 6.8 \cdot 10^4$ are consistent with all of the requirements, and form the starting point for this study. The ability to observe signatures of these models has been examined previously, both at colliders~\cite{Conley:2010du} and with astrophysical experiments~\cite{Cotta:2010ej}. Here, we turn to RAZ in chargino-neutralino production with decays to trileptons. Since a large fraction of this pMSSM model set features very small mass splittings between the $\tilde{\chi}_1^\pm$ and LSP, making the lepton from the lightest chargino decay difficult to detect, we consider only those models where the mass difference between the lightest chargino and the LSP is at least 50 GeV. In addition, we only use models expected to be observable above background with 1-10 fb$^{-1}$ at the LHC~\cite{Conley:2010du}. Finally, we show only results for models for which the production processes in our study are expected to have sufficient statistics to be detectable at an upgraded LHC with 1 ab$^{-1}$ of data at 14 TeV.

\section{Properties of Radiation Amplitude Zeros}
\label{raz}

The existence of RAZ was first discovered in the tree level calculation of $u \bar{d} \rightarrow W^+ \gamma$~\cite{Brown:1979ux}, where in the center of mass frame, the amplitude was found to vanish when the angle between the incoming $\bar{d}$ and the outgoing photon satisfied $\cos \theta^* = \frac{1}{3}$. The theory of amplitude zeros was then developed further~\cite{Zhu:1980sz,Brodsky:1982sh}, and RAZ were found to follow from the factorization of amplitudes with external gauge bosons, specifically, the ability to factor out parts of the amplitude that depend only on the gauge charges. The amplitude zero in $W \gamma$ production has now been observed experimentally, both at the Tevatron~\cite{:2008vja} and at the LHC~\cite{Chatrchyan:2011rr}.

Using the factorization property of these amplitudes, it was shown~\cite{Brodsky:1982sh} that for a process involving $n$ initial- and final-state particles, with charges $Q_i$ and 4-momenta $(p_i)$, $1 \le i \le n$, as well as a photon $\gamma$ with 4-momentum $q$, the amplitude has a radiation zero at tree level when  
\begin{equation}
\frac{Q_i}{p_i \cdot q} = \frac{Q_j}{p_j \cdot q}
\end{equation}
for all $i, j$. For a 2 $\rightarrow$ 2 scattering process $f_1 + f_2 \rightarrow X + \gamma$, this condition comprises two independent relations, with one being equivalent to charge and 4-momentum conservation. The other condition for the amplitude zero may be simply written as
\begin{equation}
\label{razpos}
\cos \theta^* = \frac{Q_{f_2} - Q_{f_1}}{Q_{f_1} + Q_{f_2}}
\end{equation}
where the $Q_{f_i}$ are the charges of the incoming particles and $\theta^*$ is the angle between the incoming $f_1$ and the outgoing photon, in the limit that all the particle masses are negligible compared to $\sqrt{s}$. At lower energies where mass effects become important, the position of the amplitude zero in $\cos \theta^*$ shifts, and the RAZ is generally more shallow in depth.

The above result can be extended to processes involving the emission of other bosons. In reactions with final-state gluons, the RAZ tend to get washed out after summing over color states~\cite{Deshpande:1994vf}. For massive gauge bosons, the amplitudes associated with their production with longitudinal polarization do not vanish, even at high energies, and so the zero is only approximate. In addition, the minimum of the angular distribution may shift considerably from the position of the original RAZ, due to the interference of the position of the true amplitude zero for transversely polarized production with the angular distribution for longitudinally polarized production. For example, at high energies, the approximate amplitude zero in $WZ$ associated production occurs at the position given by Eq.~\ref{razpos}, with the electric charges of the incoming quarks replaced by their left-handed couplings to the $Z$~\cite{Baur:1994ia}. More generally, the expression for the scattering angle where the RAZ occurs contains an overall multiplicative factor that depends on $\sqrt{s}$ and the $W$ and $Z$ masses, and approaches unity in the high energy limit.

In attempting to supersymmetrize RAZ, we first note that in exact supersymmetry, if a RAZ exists in a given SM process, an analogous zero will be found in its supersymmetric counterpart~\cite{Robinett:1984gb}. For instance, in the limit of exact supersymmetry, an amplitude zero is expected in $\tilde{W} \tilde{\gamma}$ production at $\cos \theta^* = \frac{1}{3}$, where $\theta^*$ is now the angle between the incoming $\bar{d}$ and the final-state \textit{photino}, in the CM frame~\cite{Barger:1983wc}. It should also be noted that RAZ have been predicted to occur in SUSY processes that have no SM analog, \ie, charged Higgs decay~\cite{Mukhopadhyaya:1990ba}; we do not consider this possibility here.

We now generalize the results of~\cite{Brodsky:1982sh} and~\cite{Barger:1983wc} to investigate the positions of potential amplitude zeros in chargino-neutralino associated production in the MSSM. For simplicity, let us first consider in turn the extreme cases where the produced neutralino is a pure bino, wino, or Higgsino eigenstate. If the couplings of the neutralino to the $u_A$ and $\bar{d}_B$ quarks are $g_u$ and $g_{\bar{d}}$ respectively, where $A, B = L, R$ denote the chiralities of the quarks, then we would expect an amplitude zero to exist in the parton level amplitude for $u_A \bar{d}_B \rightarrow \tilde{\chi}_i^+ \tilde{\chi}_j^0$ at
\begin{equation}
\label{susyrazpos}
\cos \theta^* = \frac{g_{\bar{d}} - g_u}{g_u + g_{\bar{d}}},
\end{equation}
in the high energy limit of negligible sparticle masses, where $\theta^*$ is the scattering angle between the $u$ and the neutralino in the CM frame. In general, we must average over all possible initial fermion chiralities. However, when the neutralino is a pure wino, only the fermions that are charged under $SU(2)_L$ may participate. Since the $u_L$ and $\bar{d}_R$ are both in $SU(2)_L$ doublets with $T_3 = +\frac{1}{2}$, they have identical couplings under the weak interaction, and thus a pure $\tilde{W}^3$ neutralino yields a RAZ at $\cos \theta^* = 0$.

On the other hand, when the neutralino is a bino, all initial fermion chiralities contribute to chargino-neutralino production. The bino has non-vanishing chiral couplings to the quarks, and we must average over all possible quark chiralities. From Eq.~\ref{susyrazpos}, we see that each set of initial chiralities gives a different RAZ position, and when all possible chiralities are combined, the zero is completely washed out. Furthermore, since the $u_L$ and $\bar{d}_R$ belong to $SU(2)_L$ doublets with opposite hypercharge, the denominator of the RHS of Eq.~\ref{susyrazpos} becomes zero for this specific set of quark chiralities.

Lastly, since light quarks have negligible couplings to the Higgs sector, there is no amplitude zero when the neutralino is a Higgsino. It is also interesting to note that because of the coupling structure of supersymmetry, a Higgsino-like neutralino cannot be produced at tree level in association with a gaugino-like chargino, or vice versa. This structure implies that the RAZ described above for chargino-neutralino production when the neutralino is wino-like is only observable if the chargino has appreciable wino content as well.

\begin{table}
\begin{center}
\caption{Predictions for amplitude zeros in associated chargino-neutralino production for the various weak eigenstates.}
\label{nxcases}
\begin{tabular}{|c|c|c|}
\hline
& $\tilde{W}^\pm$ & $\tilde{H}^\pm$ \\
\hline
$\tilde{B}$ & No physical RAZ & No tree level diagrams \\
$\tilde{W}^3$ & RAZ at $\cos \theta^* = 0$ & No tree level diagrams \\
$\tilde{H}_{1,2}$ & No tree level diagrams & No physical RAZ \\
\hline
\end{tabular}
\end{center}
\end{table}

\begin{figure}
\centering
\subfloat[]{\includegraphics[height=3in]{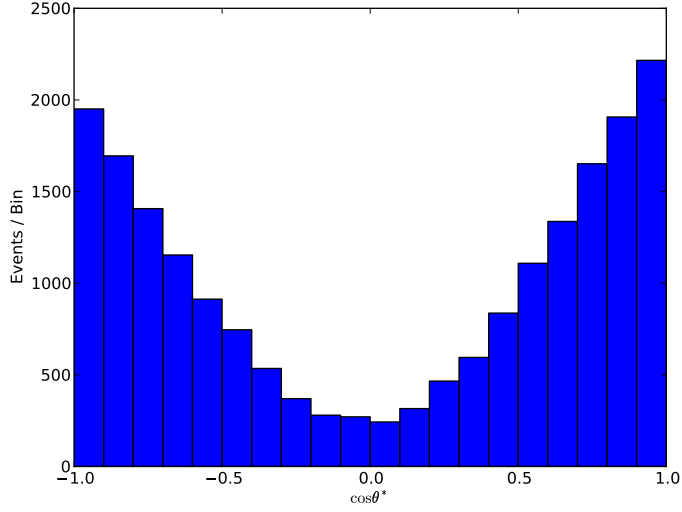}} \\
\subfloat[]{\includegraphics[height=3in]{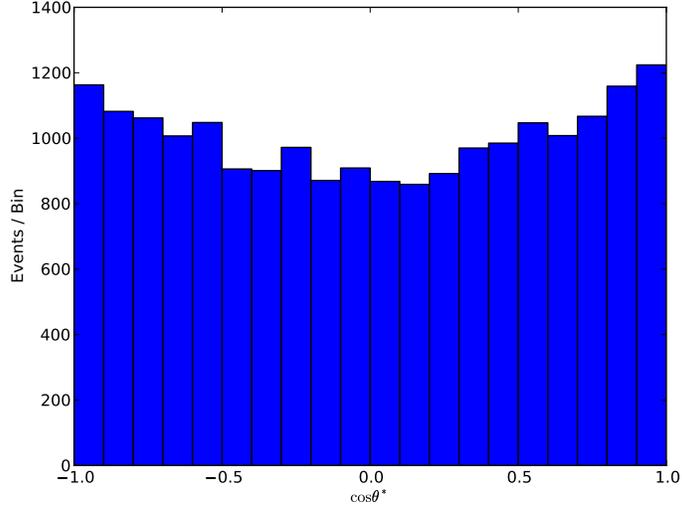}}
\vspace*{0.5cm}
\caption{The angular distribution in the center-of-mass frame for the process $u \bar{d} \rightarrow \tilde{\chi}_1^+ \tilde{\chi}_2^0$ for the benchmark point SPS4, where the neutralino is almost purely wino. Here, $\theta^*$ is the angle between the quark and the neutralino. In the top panel, the energies of the incoming partons are taken to be 1 TeV for illustration. Note the amplitude zero at $\cos \theta^* = 0$. Incorporating the full parton distribution functions at the 14 TeV LHC, the analogous distribution is shown in the bottom panel. The parton-level cross section is larger near threshold, and at lower $\sqrt{\hat{s}}$, the zero gets shallower, but does not change position.}
\label{parton}
\end{figure}

These results are summarized in Table~\ref{nxcases}. The position of the RAZ for a wino-like neutralino is particularly attractive, both because it remains unaffected by the masses of the final state sparticles, by analogy with $WZ$ production, and because it predicts that the effects of the RAZ will be seen in the central region of the detector. As an example, in Figure~\ref{parton} we show the CM frame angular distribution for $\tilde{\chi}_1^\pm \tilde{\chi}_2^0$ production for the Snowmass benchmark point SPS4~\cite{Allanach:2002nj} (based on mSUGRA), where the $\tilde{\chi}_2^0$ is almost a pure wino eigenstate, with events generated using MadGraph/MadEvent 4~\cite{Alwall:2007st}. We see the clear appearance of the RAZ in the figure at the position indicated by Table~\ref{nxcases}, with a position that is independent of CM energy.

Since chargino-neutralino production can only exhibit a RAZ when the neutralino is a wino, the presence of an amplitude zero may be used to infer the wino content of the neutralino. We now turn to practical methods of observing this amplitude zero.

\section{Observing Supersymmetric RAZ at the LHC}
\label{analysis}

As shown above, associated chargino-neutralino production has a radiation amplitude zero at $\cos \theta^* = 0$ in the high energy limit, when the neutralino is a pure wino and the chargino has a non-trivial wino component. We nos discuss how this may be observed at the LHC. The strategy of looking at the rapidity correlations of the final state particles to see the amplitude zero, as proposed for $W \gamma$ production in~\cite{Baur:1994sa}, cannot be applied here because the chargino and neutralino decays are more complicated than the decay of the $W$ and additionally involve the massive but unobservable LSP. If the neutralino is not the $\tilde{\chi}_1^0$, which we have assumed to be the LSP, it may decay to a pair of leptons and the LSP, either through a slepton cascade or a $Z^*$ + LSP decay. Similarly, the chargino may decay to a lepton, neutrino, and the LSP. Together, these two decays yield the ``golden'' trilepton signature~\cite{Baer:1994nr} that is known for being a relatively clean signal at hadron colliders and which we consider here. There is, in principle, a RAZ in chargino-LSP production when the lightest neutralino is a wino. However, because the LSP is stable, chargino-LSP production does not yield a trilepton signature, and there are not enough visible final state particles in this situation to observe the RAZ.

The RAZ in $\tilde{\chi}_1^\pm \tilde{\chi}_2^0$ production may be observed as follows. The leptons from the chargino and neutralino decays will tend to have more transverse momentum if their parent particles are produced at a large angle with respect to the beam axis. If the neutralino is a wino, the radiation amplitude zero will lead to a deficit of events at small $\cos \theta^*$, implying that there will be fewer high-$p_T$ leptons from the decaying chargino and neutralino. We can thus test whether the chargino-neutralino production amplitude zero is present, \ie, whether the neutralino has a large wino content, by comparing the leptonic $p_T$ spectrum to that of a control process that is known not to exhibit an amplitude zero. For models with a large wino content of the neutralinos, the lepton spectra will be softer and the effect of the RAZ should tend to be more pronounced.

\begin{figure}
\centerline{
\includegraphics[height=6in]{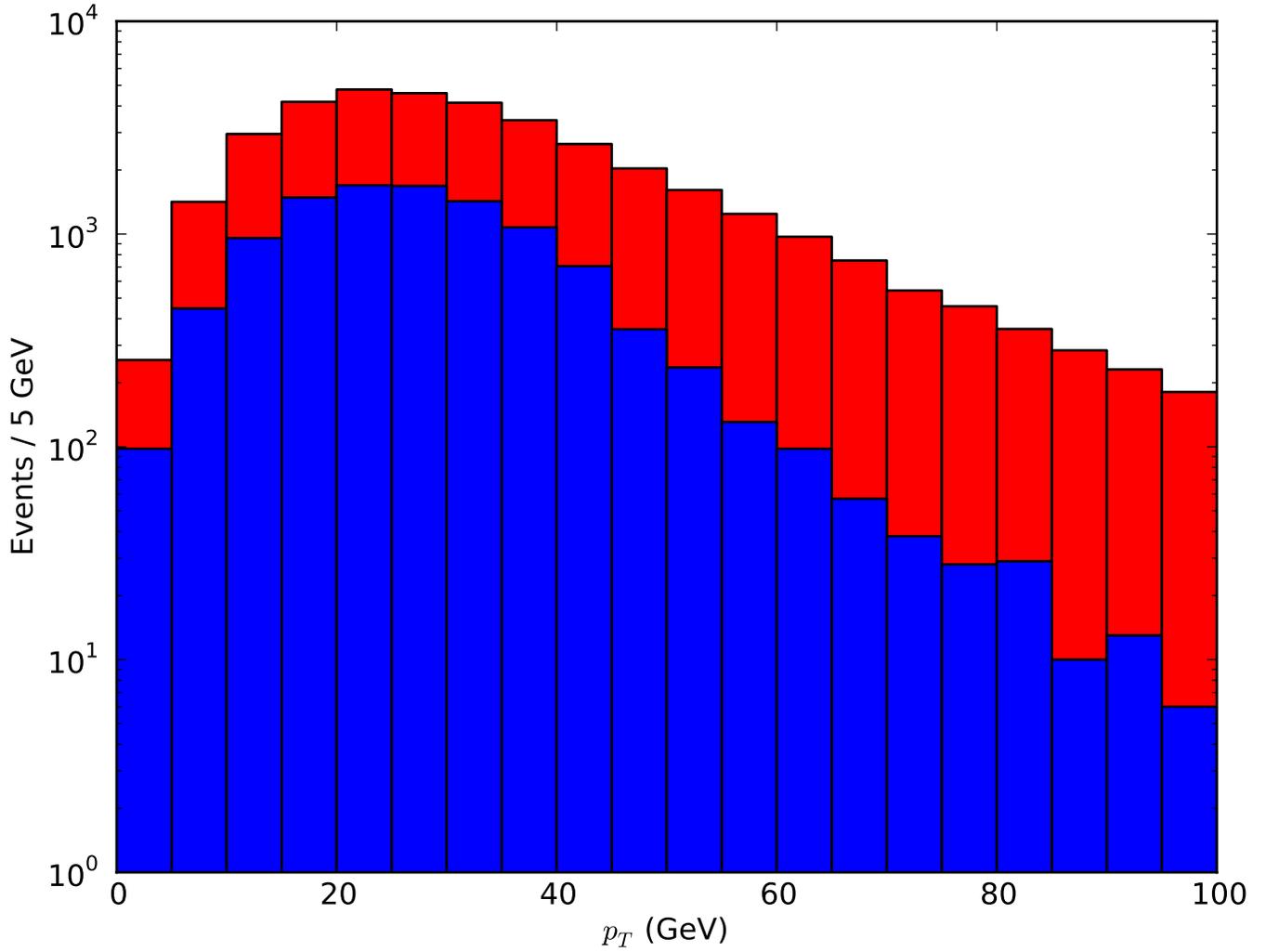}}
\vspace*{0.5cm}
\caption{The $p_T$ distribution of the softest lepton in $\tilde{\chi}_1^+ \tilde{\chi}_1^- \rightarrow 2 \ell + \mathrm{MET}$ (red) and $\tilde{\chi}_1^\pm \tilde{\chi}_2^0 \rightarrow 3 \ell + \mathrm{MET}$ (blue) events for the SUSY benchmark point SPS4, at $\sqrt{s} = 14 \mbox{ TeV}$. Because of the amplitude zero in the latter process, the spectrum falls off more sharply.}
\label{sps4pt}
\end{figure}

In Figure~\ref{sps4pt}, we show the $p_T$ distribution of the softest lepton for $\tilde{\chi}_1^\pm \tilde{\chi}_2^0 \rightarrow 3 \ell + \mathrm{MET}$ production for the SUSY benchmark point SPS4~\cite{Allanach:2002nj}. For comparison, we also show the corresponding $p_T$ spectrum of the softest lepton in a control process, $\tilde{\chi}_1^+ \tilde{\chi}_1^- \rightarrow 2 \ell + \mathrm{MET}$, which does not have a radiation amplitude zero. Note that $\tilde{\chi}_2^0 \approx \tilde{W^3}$ for SPS4, and so we see that indeed the chargino-neutralino softest lepton $p_T$ distribution falls off significantly faster than for its chargino pair production counterpart, displaying the presence of an amplitude zero. For all of the simulations in this analysis, events were generated for the LHC at 14 TeV using PYTHIA 6.4~\cite{Sjostrand:2006za}, with initial-state radiation, final-state radiation, multiple interactions, and fragmentation turned off. Because we only consider signatures with leptons and missing energy in the final state, we do not expect these choices to significantly affect our results. Throughout, we require that leptons be observed within $|\eta| < 2.5$.

\begin{figure}
\centering
\subfloat[]{\includegraphics[height=3in]{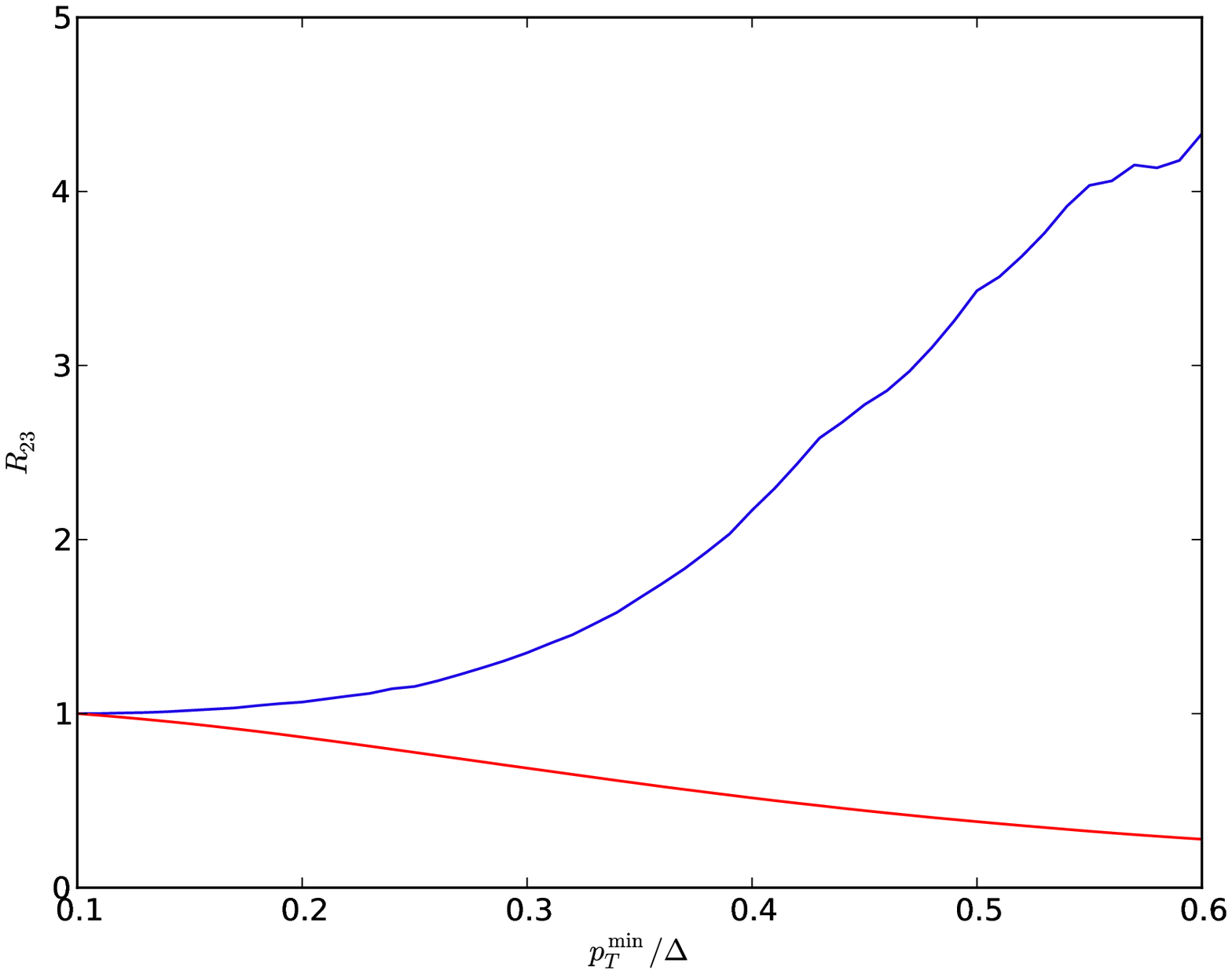}} \\
\subfloat[]{\includegraphics[height=3in]{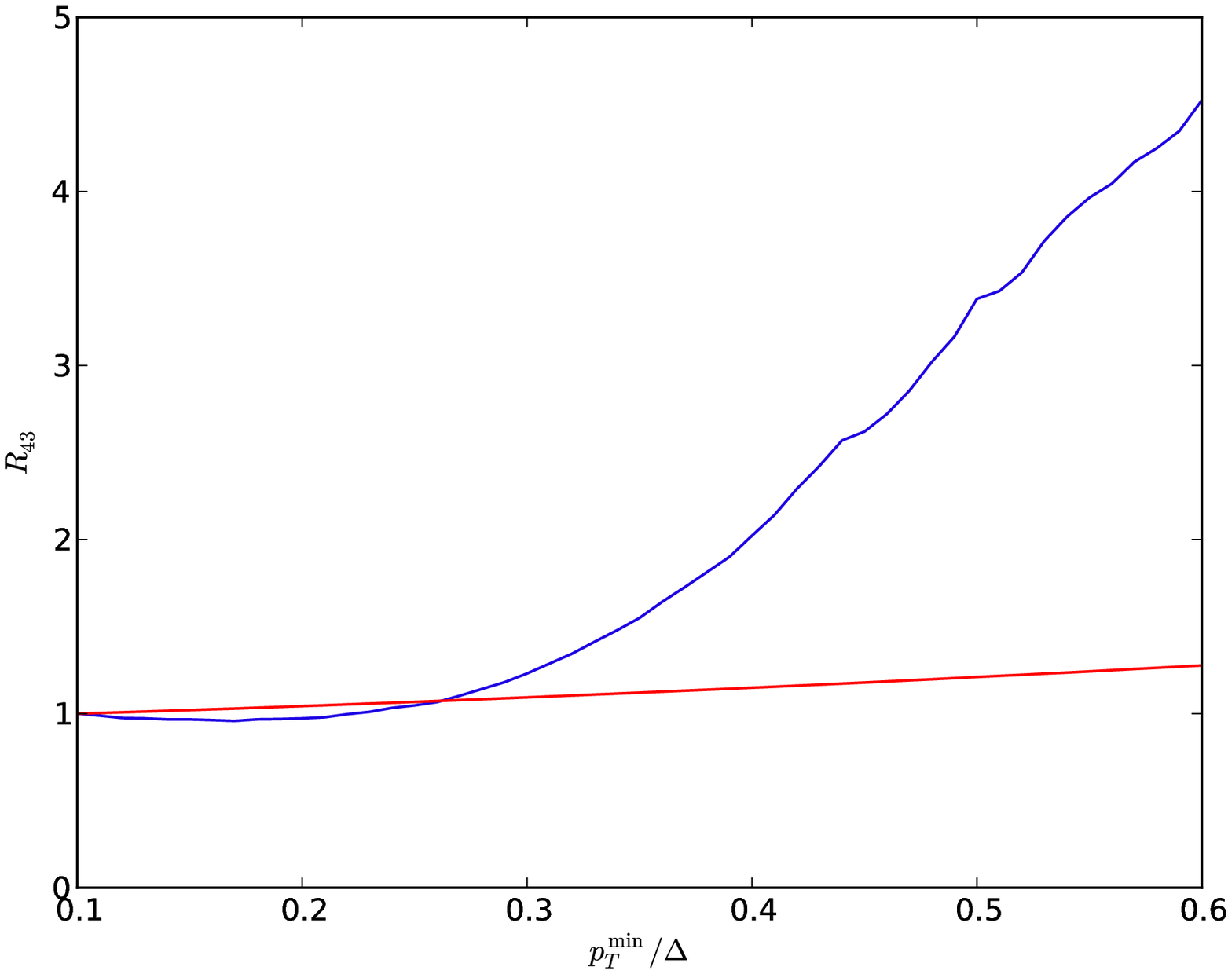}}
\vspace*{0.5cm}
\caption{The ratios $R_{23}$ (top) and $R_{43}$ (bottom), of $2 \ell + \mbox{MET}$ to $3 \ell + \mbox{MET}$ and $4 \ell + \mbox{MET}$ to $3 \ell + \mbox{MET}$ events respectively, as a function of the lepton $p_T$ cut, for the SUSY benchmark points SPS4 (blue) and SPS9 (red).}
\label{spsplot}
\end{figure}

Following a procedure similar to that presented in~\cite{Baur:1993ix}, we attempt to ascertain the presence of a radiation amplitude zero in chargino-neutralino production by examining the ratio of number of $\tilde{\chi}_1^+ \tilde{\chi}_1^- \rightarrow 2 \ell + \mathrm{MET}$ events to that of $\tilde{\chi}_1^\pm \tilde{\chi}_2^0 \rightarrow 3 \ell + \mathrm{MET}$, as a function of the softest lepton $p_T$ cut. This ratio is straightforward to measure experimentally, and we expect it to rise significantly with the $p_T$ cut when there is an amplitude zero. Specifically, we define
\begin{align}
R_{23}(p_T^\mathrm{min}) &= C_{23} \cdot \frac{N(\tilde{\chi}_1^+ \tilde{\chi}_1^- \rightarrow 2 \ell + \mbox{MET})}{N(\tilde{\chi}_1^\pm \tilde{\chi}_2^0 \rightarrow 3 \ell + \mbox{MET})},
\end{align}
where $C_{23}$ is a normalization factor chosen such that $R_{23}(0.1 \Delta) = 1$, with $\Delta = m(\tilde{\chi}_1^\pm) - m(\tilde{\chi}_1^0)$. We are interested in how $R_{23}$ changes as the softest lepton $p_T$ cut is increased relative to the characteristic scale of the momentum of the lepton from $\tilde{\chi}_1^\pm$ decay, which is roughly $\Delta$. Figure~\ref{spsplot} shows this ratio for two different SPS benchmark points, for illustration. For SPS4, with $\tilde{\chi}_2^0 \approx \tilde{W}^3$, as the softest lepton $p_T$ cut increases, there are comparatively fewer events left from the chargino-neutralino production process exhibiting an amplitude zero, and so the ratio rises. Conversely, the $\tilde{\chi}_2^0$ in SPS9 is nearly pure bino, and $\tilde{\chi}_1^\pm \tilde{\chi}_2^0$ production does not feature an amplitude zero in this scenario; and so here, we find that $R_{23}$ decreases as the $p_T$ cut is increased. This demonstrates the promise of this approach to determine the wino content of neutralinos.

If the leptons from the neutralino decay are significantly less energetic than those from the chargino decay, the softest lepton $p_T$ spectra for the $\tilde{\chi}_1^+  \tilde{\chi}_1^- \rightarrow 2 \ell + \mathrm{MET}$ and $\tilde{\chi}_1^\pm \tilde{\chi}_2^0 \rightarrow 3 \ell + \mathrm{MET}$ processes will not be directly comparable. Such a situation may arise, for instance, if both the $\tilde{\chi}_1^\pm$ and $\tilde{\chi}_2^0$ decay through cascades, and the characteristic momentum scales for the leptons from the chargino and neutralino decays are $m(\tilde{\chi}_1^\pm) - m(\tilde{\nu}_\ell) \gg m(\tilde{\chi}_2^0) - m(\tilde{\ell})$. In this case, we can instead compare the softest lepton $p_T$ of the processes $\tilde{\chi}_2^0 \tilde{\chi}_2^0 \rightarrow 4 \ell + \mathrm{MET}$ and $\tilde{\chi}_1^\pm \tilde{\chi}_2^0 \rightarrow 3 \ell + \mathrm{MET}$. While the four lepton signature tends to have lower statistics, since we typically have $m(\tilde{\chi}_2^0) > m(\tilde{\chi}_1^\pm)$, it complements our first approach above. We may thus also construct the additional ratio
\begin{align}
R_{43}(p_T^\mathrm{min}) &= C_{43} \cdot \frac{N(\tilde{\chi}_2^0 \tilde{\chi}_2^0 \rightarrow 4 \ell + \mbox{MET})}{N(\tilde{\chi}_1^\pm \tilde{\chi}_2^0 \rightarrow 3 \ell + \mbox{MET})}
\end{align}
as a function of the lepton transverse momentum cut. As before, $C_{43}$ is chosen such that $R_{43}(0.1 \Delta) = 1$. Figure~\ref{spsplot} shows $R_{43}$ for the SUSY benchmark points SPS4 and SPS9, again demonstrating the ability of our observables to investigate neutralino mixing.

\begin{figure}
\centerline{
\includegraphics[height=6in]{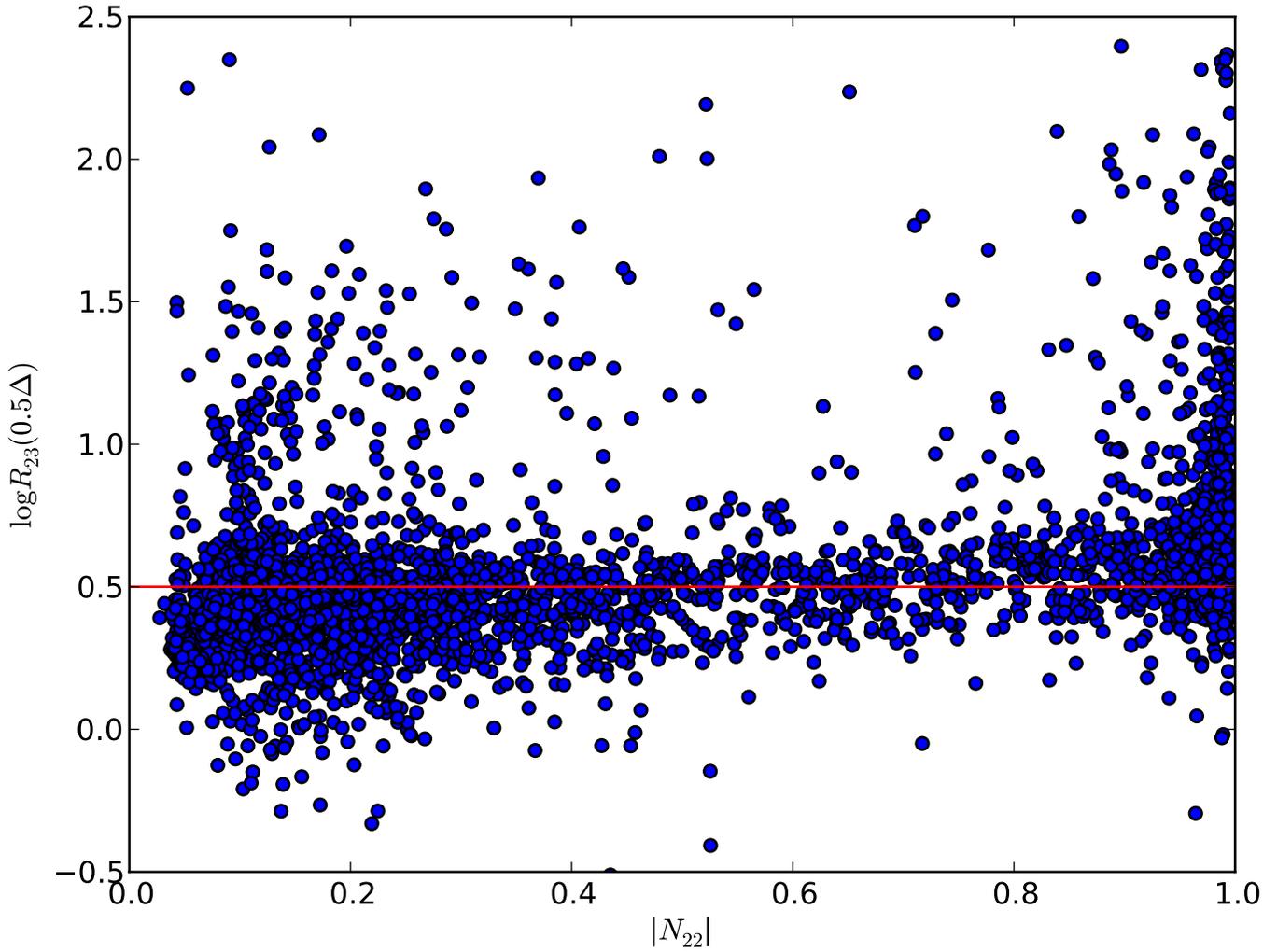}}
\vspace*{0.5cm}
\caption{$R_{23}(0.5 \Delta)$ for all models in our set with sufficient statistics. Since $R_{23}$ is normalized such that $R_{23}(0.1 \Delta) = 1$, this measures how much $R_{23}$ changes with the lepton $p_T$ cut. For models with low values of $R_{23}(0.5 \Delta)$, \eg, those below the sample red line, the second neutralino tends to have low wino content.}
\label{r23n22}
\end{figure}

\begin{figure}
\centerline{
\includegraphics[height=6in]{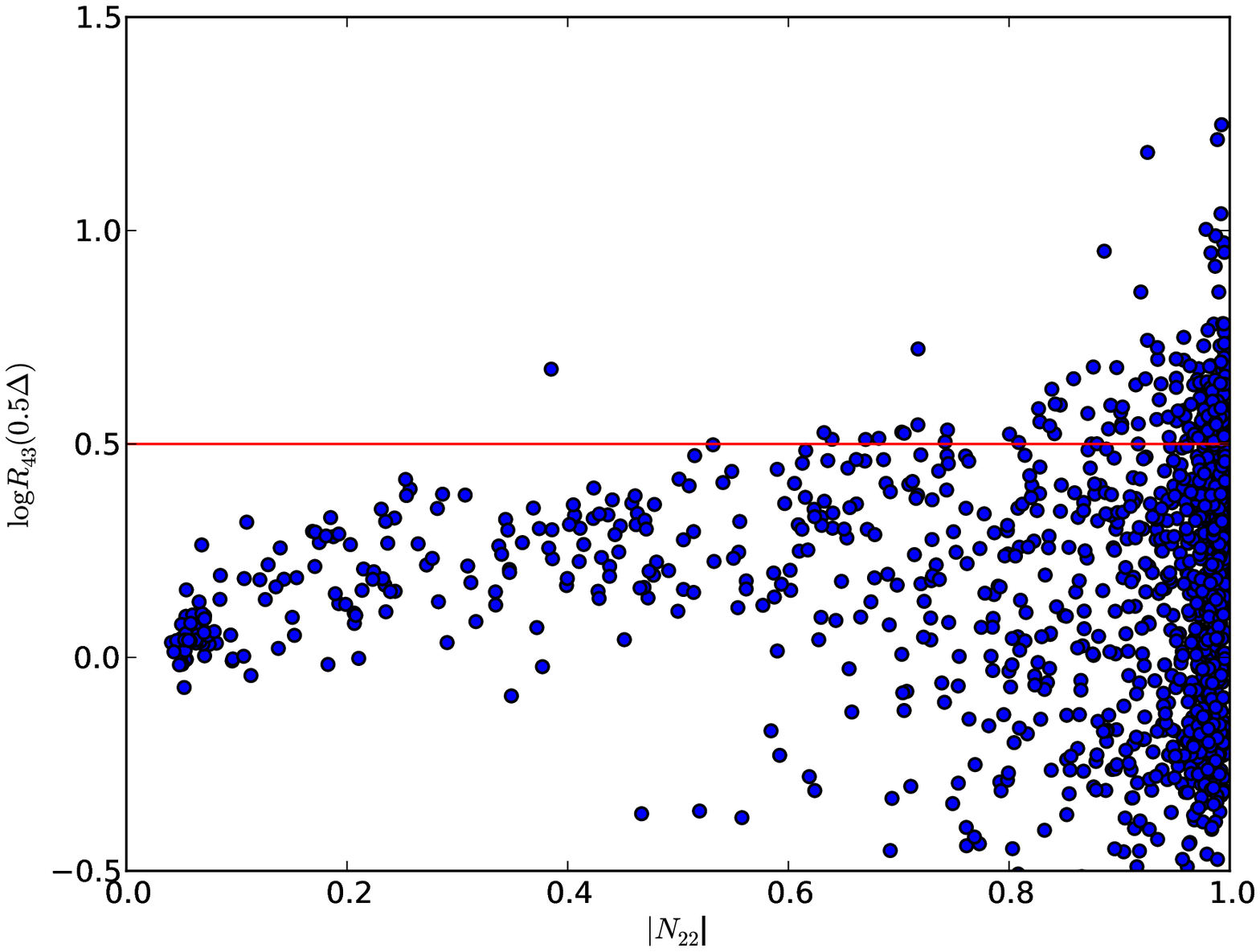}}
\vspace*{0.5cm}
\caption{$R_{43}(0.5 \Delta)$ for all models in our set with sufficient statistics. Since $R_{43}$ is normalized such that $R_{43}(0.1 \Delta) = 1$, this measures how much $R_{43}$ changes with the lepton $p_T$ cut. For models with high values of $R_{43}(0.5 \Delta)$, \eg, those above the sample red line, the second neutralino tends to have high wino content.}
\label{r43n22}
\end{figure}

We now examine these two ratios in our pMSSM model set. Many models feature very small $\tilde{\chi}_1^\pm$-$\tilde{\chi}_1^0$ mass splittings, which would render the softest lepton unobservable for processes involving leptonic chargino decays. We have thus restricted ourselves to the subset of models where $\Delta > 50 \mbox{ GeV}$, corresponding to approximately 16\% of the full model set. For this smaller set of models, we generated $\tilde{\chi}_1^+ \tilde{\chi}_1^-$, $\tilde{\chi}_1^\pm \tilde{\chi}_2^0$, and $\tilde{\chi}_2^0 \tilde{\chi}_2^0$ events, and constructed the ratios $R_{23}$ and $R_{43}$. When the produced gauginos/Higgsinos have small leptonic branching ratios or are very heavy, the statistics are generally low. In our work, we have further restricted ourselves to those models where the 14 TeV LHC with 1 ab$^{-1}$ of data could acquire enough statistics to yield meaningful values for the ratios $R_{23}$ and $R_{43}$ through a minimum $p_T$ cut on the softest lepton of $0.5 \Delta$. Figures~\ref{r23n22} and~\ref{r43n22} show how $R_{23}$ and $R_{43}$ vary with $|N_{22}|$, the magnitude of the neutralino mixing matrix element corresponding to the wino content of the second neutralino. These figures show the results for approximately $4.4 \cdot 10^3$ and $1.2 \cdot 10^3$ models, respectively. The trend in these figures is generally as expected, with higher values of $|N_{22}|$ corresponding to larger ratios. As we increase the softest lepton $p_T$ cut, the overall number of events for our models drop, but the rising trend of the ratios $R_{23}$ and $R_{43}$ becomes easier to discern. Again, we have considered the values of the ratios at a lepton $p_T$ cut of $0.5 \Delta$ to strike a balance between statistics and clarity.

While the models with low values of $R_{23}(0.5 \Delta)$ indeed have generally non-wino-like $\tilde{\chi}_2^0$, there are outliers in the upper region of Figure~\ref{r23n22}, with $R_{23}$ rising faster than expected. The converse is true in Figure~\ref{r43n22}, where the region corresponding to high $R_{43}$ is very clean. In addition to the issues related to the spectrum dependence discussed above, another reason for these features is the number of unobservable leptons in our processes. There are 4 leptons (charged and neutral) in the final state of each reaction we have considered, but either 0, 1, or 2 neutrinos. The ratios $R_{23}$ and $R_{43}$ depend on the numbers of events from these processes which survive a $p_T$ cut on the softest charged lepton. Since this cut does not remove events with low $p_T$ neutrinos, a given event with four total charged leptons and missing energy from the LSP in the final state is more likely to pass such a cut tnan if one or more of the final state leptons are unobservable neutrinos, as there is then more accepted phase space in the sparticle decays.

\begin{figure}
\centerline{
\includegraphics[height=6in]{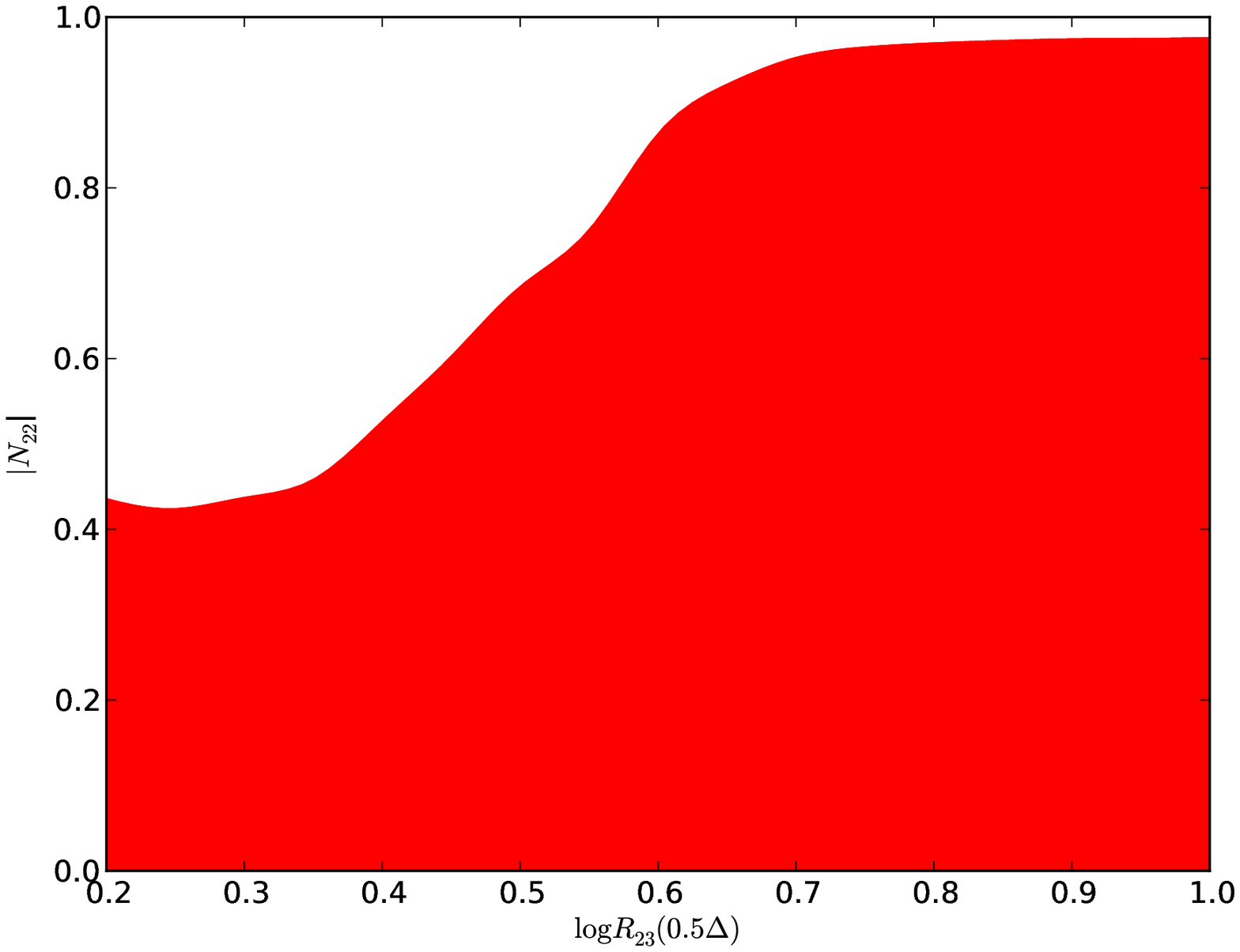}}
\vspace*{0.5cm}
\caption{Upper limit on $|N_{22}|$ from $R_{23}$ in our pMSSM model set. The uncolored region is excluded with 90\% confidence. Note that because of outliers in Figure~\ref{r23n22}, the lower limit on $|N_{22}|$ from $R_{23}$ is not useful.}
\label{limitr23}
\end{figure}

\begin{figure}
\centerline{
\includegraphics[height=6in]{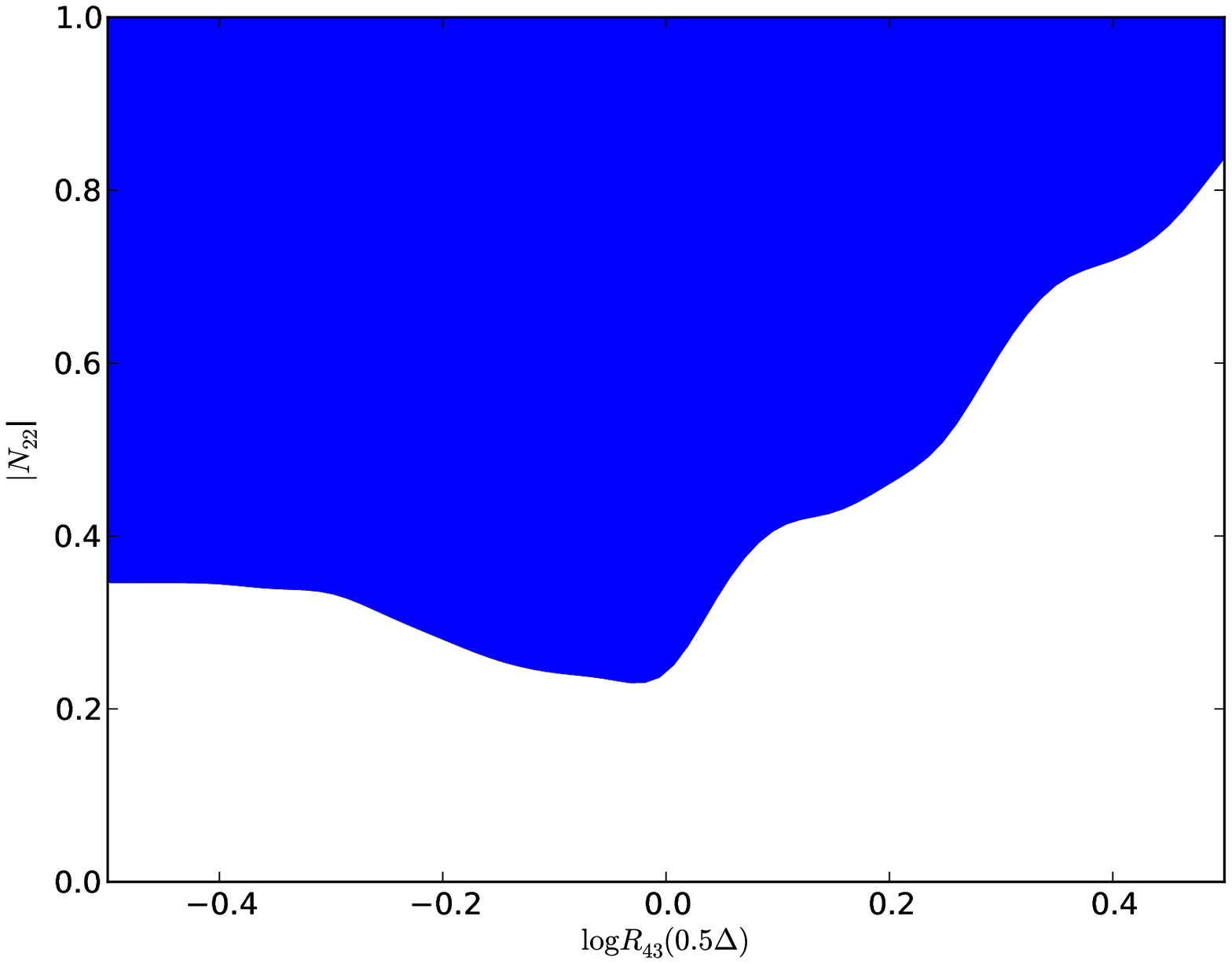}}
\vspace*{0.5cm}
\caption{Lower limit on $|N_{22}|$ from $R_{43}$ in our pMSSM model set. The uncolored region is excluded with 90\% confidence. Note that because of outliers in Figure~\ref{r43n22}, the upper limit on $|N_{22}|$ from $R_{43}$ is not useful.}
\label{limitr43}
\end{figure}

Finally, Figures~\ref{limitr23} and~\ref{limitr43} show our 90\% confidence level upper and lower limits on $|N_{22}|$ from the smaller pMSSM model set employed in the previous figures as functions of $R_{23}(0.5 \Delta)$ and $R_{43}(0.5 \Delta)$, respectively. We see that taken together, these two measurements can provide meaningful constraints. With enough integrated luminosity, measuring these ratios at the LHC should provide a simple way to bound the wino content of the neutralino. Even more data would enable the construction of $R_{23}$ and $R_{43}$ with a higher leptonic transverse momentum cut $p_T^\mathrm{min}$, potentially yielding a significantly better constraint on the neutralino's wino content.

\section{Conclusion}
\label{conclusion}

We have investigated the potential of the LHC to determine the wino content of the neutralino present in supersymmetric theories through the measurement of radiation amplitude zeros, a feature of certain processes first discovered in the early days of the Standard Model. Looking at chargino-neutralino associated production is the natural extension to supersymmetry of studies of $W \gamma$ and $WZ$ production, and we expect analogous amplitude zeros if the neutralino is a wino eigenstate. Even though the dramatic RAZ signature is somewhat masked at a hadron collider such as the LHC, we have shown that it should still be visible in many cases, using an appropriate observable.

Using a model set derived from a scan of the 19-dimensional pMSSM parameter space, we have demonstrated the ability of this approach to place bounds on the wino content of $\tilde{\chi}_2^0$. To demonstrate, we used PYTHIA to compare the event rate for chargino-neutralino production to that of control processes which never exhibit a RAZ, as a function of the transverse momentum cut on the softest lepton. We observe that the number of chargino-neutralino events tends to fall off faster with increasing lepton $p_T$ cut when the neutralino is a wino. We find that by constructing different observables, both upper and lower limits on the magnitude of the neutralino mixing matrix element $|N_{22}|$ may be obtained with minimal dependence on the exact spectrum of the theory.

While we have only considered production of the second neutralino $\tilde{\chi}_2^0$, our analysis will also apply to the heavier neutralinos as well. In theory, similar ratios of events passing minimum $p_T$ cuts could be used to constrain the mixing matrix elements $|N_{32}|$ and $|N_{42}|$. However, to construct these ratios, a significant sample of events would be needed. Depending on the SUSY spectrum realized by nature, particularly the masses and branching ratios of the heavier neutralinos, obtaining bounds on the content of heavier neutralinos may or may not be beyond the scope of the LHC.

It should be noted that if a specific model of SUSY breaking is realized, our results remain relevant. Since there are fewer parameters in these scenarios, it is possible that the neutralino mixing matrix may be implicitly determined in the early years of the LHC, through the extraction of the model parameters from other processes; for instance, see~\cite{Hinchliffe:1996iu} in the case of mSUGRA. Indeed, global fits have been attempted in more general MSSM scenarios as well~\cite{Lafaye:2004cn}, and our analysis complements these works. In such cases, we would already know the mixing in the neutralino sector at some level, without having to wait for sufficiently large samples of chargino-neutralino events to perform our RAZ analysis. However, the predicted RAZ behavior of the various chargino-neutralino production processes could then be used to verify the coupling structure of the newly discovered supersymmetric particles, just as studies of $W \gamma$ continue to test the couplings of the $W$ and the gauge structure of the Standard Model.

Radiation amplitude zeros have provided a powerful way to probe the Standard Model in the past, and as the LHC is on the verge of determining the mechanism underlying electroweak symmetry breaking, we have shown how an old technique can be applied to a new theory. In the coming years, we expect that amplitude zeros will be very useful to investigate the structure of supersymmetry and beyond.

\section*{Acknowledgements}

The authors are grateful for discussions with S. Brodsky, J. Conley, R. Cotta, J. Gainer, and M.P. Le. AI was supported in part by the Natural Sciences and Engineering Research Council of Canada.

\newpage

%
%%%%%%%%%%%%%%%%%%--- References
%%%%%%%%%%%%%%%%%%%%%%%%%%%%%%%%%%%%%%%%%%%%%%%%%%%%%%%
\def\IJMP #1 #2 #3 {Int. J. Mod. Phys. A {\bf#1},\ #2 (#3)}
\def\MPL #1 #2 #3 {Mod. Phys. Lett. A {\bf#1},\ #2 (#3)}
\def\NPB #1 #2 #3 {Nucl. Phys. {\bf#1},\ #2 (#3)}
\def\PLBold #1 #2 #3 {Phys. Lett. {\bf#1},\ #2 (#3)}
\def\PLB #1 #2 #3 {Phys. Lett. B {\bf#1},\ #2 (#3)}
\def\PR #1 #2 #3 {Phys. Rep. {\bf#1},\ #2 (#3)}
\def\PRD #1 #2 #3 {Phys. Rev. D {\bf#1},\ #2 (#3)}
\def\PRL #1 #2 #3 {Phys. Rev. Lett. {\bf#1},\ #2 (#3)}
\def\PTT #1 #2 #3 {Prog. Theor. Phys. {\bf#1},\ #2 (#3)}
\def\RMP #1 #2 #3 {Rev. Mod. Phys. {\bf#1},\ #2 (#3)}
\def\ZPC #1 #2 #3 {Z. Phys. C {\bf#1},\ #2 (#3)}

\end{document}